\documentclass[]{aa}
\usepackage{graphicx}
\usepackage{txfonts}

\begin{document}

\title{Oxygen-rich disk in the V778 Cyg system resolved.\thanks{Table 2 is only available in
electronic form at {\tt http://wwww.edpscience.org}}}

\subtitle{}

\author{R. Szczerba\inst{1}
        \and M. Szymczak\inst{2}
        \and N. Babkovskaia\inst{3}
        \and J. Poutanen\inst{3}
        \and A.M.S. Richards\inst{4}
        \and M.A.T. Groenewegen\inst{5}}

\offprints{R. Szczerba: \email{szczerba@ncac.torun.pl}}

\institute{N. Copernicus Astronomical Center, Rabia\'{n}ska 8,
               87-100 Toru\'n, Poland
  \and  Toru\'n Centre for Astronomy, Nicolaus Copernicus
              University, Gagarina 11, 87-100 Toru\'n, Poland
  \and  Astronomy Division, P.O.Box 3000, University of Oulu, FIN-90014
Oulu, Finland
  \and Jodrell Bank Observatory, University of Manchester, Macclesfield,
Cheshire SK11 9Dl, UK
   \and Instituut voor Sterrenkunde, PACS-ICC, Celestijnenlaan 200B, B-3001
Leuven, Belgium}

\date{Received  / Accepted }

\abstract
{}
{Various scenarios have been proposed to explain the presence
of silicate features associated with carbon stars, such as
\object{V778\,Cyg}.  We have attempted to constrain these theories by
means of mapping water maser mission from V778\,Cyg.}
{The 22\,GHz water maser emission from this star has been mapped using MERLIN with an 
astrometric accuracy of 25\,mas.}
{The spatially- and kinematically-resolved maser complex is displaced
by $\sim$190\,mas from the position of the C-star as measured 10 years
earlier using Tycho. Our simulations and analysis of available data
show that this position difference is unlikely to be due to proper
motion if V778\,Cyg is at the assumed distance of 1.4\,kpc. The maser
components seem to form a distorted S-shaped structure extended over
$\sim$18\,mas with a clear velocity gradient. We propose a model which
explains the observed water maser structure as an O-rich warped disk
around a companion of the C-star in V\,778 binary system, which is
seen almost edge-on.}
{Analysis of observational data, especially those obtained with
MERLIN, suggests that V778\,Cyg (and, by implication, other silicate
carbon stars) are binary systems composed of a C-rich star and a
companion which stores circumstellar O-rich material.}


          \titlerunning{O-rich disk in V778 Cyg}
          \authorrunning{Szczerba et al.}

\maketitle
%

\section{Introduction}

The silicate emission features at about 10 and 18\,$\mu$m are characteristic
of O-rich dust envelopes. Surprisingly, these features were also discovered
in the IRAS LRS data for some optically classified carbon
stars (Little-Marenin\,\cite{LM86}; Willems \& de Jong\,\cite{WdJ86}), later
termed silicate carbon stars. The detection of silicate emission from these
stars suggests that their relatively close surroundings contain oxygen-based 
dust, in spite of their photospheric chemical composition which shows
C/O\,$>$\,1. An additional argument for the persistence of O-rich material
comes from the detection of water and OH maser lines towards some silicate
carbon stars (e.g. Little-Marenin et al.\,\cite{LS94};
Engels\,\cite{engels94}; Little-Marenin et al.\,\cite{littlemareninetal88}).

Little-Marenin\,(\cite{LM86}) proposed that silicate carbon stars are
binaries consisting of  C-rich and O-rich giants, but extensive
observations (e.g. Lambert et al.\,\cite{LH90};
Engels \& Leinert\,\cite{engelsandleinert94}) did not show any evidence for
an O-rich giant in these sources.
Willems \& de Jong\,(\cite{WdJ86}) proposed that silicate carbon stars were
formed very recently due to a thermal pulse which changed the chemistry of the
star from O- to C-rich and that the O-rich material ejected before and
during the thermal pulse gives the observed silicate features.
However, the Infrared Space Observatory (ISO) spectra showed that
during the 14-year time interval between IRAS and ISO missions the shape
and intensity of the silicate features in V778\,Cyg did not change at all
(Yamamura et al.\,\cite{yamamuraetal00}). 
This suggests that O-rich material is located in some stable
configuration and that the model of a fast transition from O- to  C-rich star
cannot apply, as the silicate features should diminish in strength quite
rapidly. Presently, the most widely accepted scenario which is able to explain
this phenomenon is a binary system composed of a C-star and an unseen, most
likely main-sequence, companion with a reservoir of O-rich material
(Morris et al.\,\cite{MG87}; Lloyd Evans \,\cite{LE90}). Yamamura et
al.\,(\cite{yamamuraetal00}) argued that the dust responsible for the
observed silicate features is stored in a disk around the companion.

To test these hypotheses we observed water masers towards
V778\,Cyg at high angular resolution.  Previous observations only revealed
 unresolved radio emission  within 0\farcs5 of
the position of the optical star (Deguchi et al.\,\cite{deguchi88},
Colomer et al.\,\cite{colomer00}). We also present a simple
quantitative interpretation of the observed structure based on a
Keplerian disk model.

\section{Observations and data reduction}

The observations were taken on 2001 October 12/13 under good weather
conditions, using five telescopes of MERLIN (Diamond et
al.\,\cite{diamond03}). The longest MERLIN baseline of 217\,km gave a
fringe spacing of 12\,mas at 22\,GHz.  A bandwidth of 2\,MHz was used
divided into 256 spectral channels per baseline providing a channel
separation of 0.105\,km\,s$^{-1}$.  The velocities ($V_{\rm LSR}$)
were measured with respect to the local standard of rest. The
continuum calibrator sources were observed in 16\,MHz bandwidth. We
used the phase referencing method; 4\,min scans on V778\,Cyg were
interleaved with 2\,min scans on the source 2021+614 (at 3\fdg8 from
the target) over 11.5\,h.  VLBA observations at 8.6\,GHz resolve
2021$+$614 into two components separated by 7\,mas along the position
angle of 33\degr\ (Fey et al.\,\cite{fey96}). We detect only one
unresolved source at 22 GHz. Its absolute position coincides within less than
2\,mas with the VLBA position of the stronger component at 8.6\,GHz.
The flux density of 2021+614 of 1.48\,Jy was derived from 4C39.25,
which had a flux density of 7.5$\pm$0.3\,Jy at the epoch of our
observations (Terasranta 2002, private communication).  This source
was also used for bandpass calibration.

After initial calibration with MERLIN software, the data were processed
using the AIPS package (Greisen\,\cite{greisen94}). To derive phase
and amplitude corrections for atmospheric and instrumental effects the phase
reference source was mapped and self-calibrated. These corrections were applied
to the V778\,Cyg visibility data. The absolute position of the brightest
feature at $-$15.1\,km\,s$^{-1}$ was determined before further calibration. 
Finally, the clean components of this
image were used as a model for phase self-calibration of this channel
and the solutions were applied to all channels. Each channel
was then mapped and cleaned using a 12\,mas circular
restoring beam. We present results for total intensity (Stokes $I$)
images. The map noise of $\sim$27\,mJy\,beam$^{-1}$ in a line-free
channel was close to the predicted thermal noise level.

To determine the position and the brightness of the maser
components two dimensional Gaussian components were fitted to the
emission in each channel maps. The position uncertainty 
depends on the channel signal to noise ratio (Condon et
al.\,\cite{condon98}, Richards et al.\,\cite{richardsetal99} and references 
therein) and is lower than 1\,mas for about 80\% of the
maser components towards V778\,Cyg. The absolute position of the phase
reference source is known within $\sim$3\,mas. The
uncertainties in the absolute positions of the maser components are
dominated by errors in the telescope positions and tropospheric effects.  
Uncertainties in telescope positions of 1$-$2\,cm cause a maser position error 
of $\sim$10\,mas. The latter uncertainty was estimated by observing the
phase rate on the point source 4C39.25 which appeared to introduce a position 
error of $\la$9\,mas. We checked this using reverse
phase referencing. Emission from 15 channels around the
reference feature at $-$15.1\,km\,s$^{-1}$ was averaged and mapped.
The map obtained was used as a model to self-calibrate the raw target
data and  these solutions were then applied to the raw data of
2021+614.  The position of the reference source was shifted by only
$\sim$2\,mas with respect to the catalog position. 
These factors imply that the absolute
position accuracy of the maser source is $\la$25\,mas.
All MERLIN coordinates are given in the ICRS system.

\section{Results and discussion}
\begin{figure}[]
\resizebox{\hsize}{!}{\includegraphics{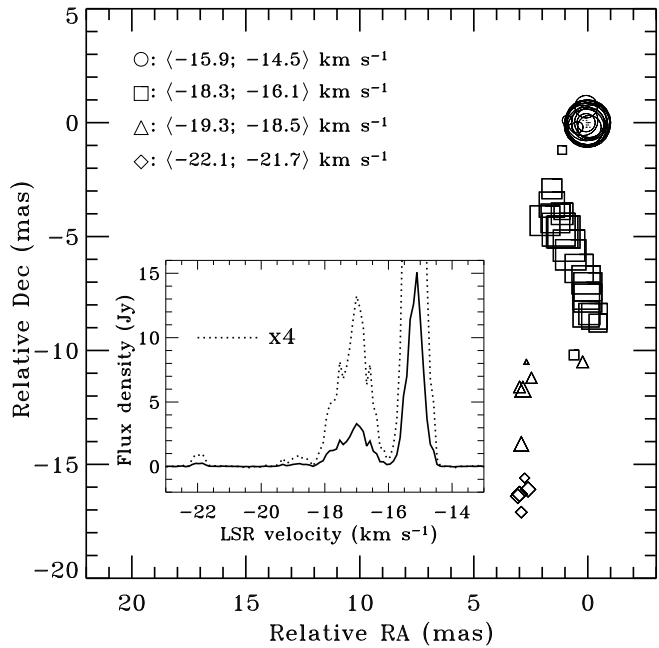}}
\caption{Positions of the water maser components in V778\,Cyg relative to the
reference component at $-$15.1\,km\,s$^{-1}$. The symbols correspond to the
velocity ranges given in the upper left corner. The size of each symbol is
proportional to the logarithm of peak brightness of the corresponding
component. {\it Inset:} MERLIN spectrum of the H$_2$O maser emission in
V778\,Cyg. The dotted line shows the spectrum magnified by a
factor of four to enlarge the weak features.}
\label{Fig1}
\end{figure}

A single, unresolved maser component brighter than
150 mJy\,beam$^{-1}$ ($\sim5\sigma$) was found in each of 51 spectral
channels.  The overall distribution of the H$_2$O maser components in
V778\,Cyg is shown in Fig.\ref{Fig1} (parameters of maser components
are listed in Table\,\ref{tab2}). The total angular extent of the
maser emission is about 18\,mas. All observed maser components
seem to form a distorted S-like shape at a position angle (P.A.) of about
$-$10\degr. However, the most spatially extended series of components 
($V_{\rm LSR}\sim$ $-$17\,km\,s$^{-1}$) are aligned along
P.A.$\approx$$+$18\degr. There is a clear velocity gradient along
the whole structure, blue-shifted in the south with respect
to the brightest northern components.

The inset in Fig.\ref{Fig1} shows the cross-correlation water maser
spectrum towards V778\,Cyg. The emission was dominated by a
$-$15\,km\,s$^{-1}$ feature. Weak emission of about 200$-$250\,mJy was
seen at $-$19 and $-$22\,km\,s$^{-1}$. The spectral shape is
roughly similar to that of the single dish spectra observed by Engels
\& Leinert\,(\cite{engelsandleinert94}) and by Nakada et
al.\,(\cite{nakada87}), apart from a $-22$\,km\,s$^{-1}$ feature
first detected on 11.3.95 by D. 
Engels\footnote{www.hs.uni-hamburg.de/DE/Ins/Per/Engels/engels/wcatalog.html}.

\begin{table}[]{}
\caption[]{Radio and optical coordinates (with errors) for V778\,Cyg.}
\label{tab1}
\centering
\begin{tabular}{l@{}l@{}l@{}l@{}}
\hline
 & Year & RA(J2000) & Dec(J2000) \\
\hline
MERLIN~~ & 2001~ & 20$^{\rm h}$36$^{\rm m}$07\fs3833\,($\pm$0\fs0008)~~~ & 60\degr05\arcmin26\farcs024\,($\pm$0\farcs025)\\
Tycho2~ & 1991~ & 20$^{\rm h}$36$^{\rm m}$07\fs4022\,($\pm$0\fs0028)~~~ & 60\degr05\arcmin26\farcs154\,($\pm$0\farcs040) \\
\hline
\end{tabular}
\end{table}

The absolute position 
of the reference feature at $-$15.1\,km\,s$^{-1}$ is given in
Table\,\ref{tab1} together with the optical position from the Tycho2
catalog (Hog et al.\,\cite{hog00}). The source is not listed in
the Hipparcos catalog. The Tycho2 positions are based on the
same observations as the Tycho1 catalog (ESA, 1997) collected by the
star mapper of the ESA Hipparcos satellite, but Tycho2 is much bigger
and more precise, owing to a more advanced reduction technique. The mean 
satellite observation epoch is 1991.5. The
coordinates given in the Tycho2 catalog are in the ICRS system and for
a star of VT\,=\,10.5 such as V778\,Cyg, the estimated astrometric
error is at a level of 40\,mas. For about 100\,000 stars (among them for 
V778\,Cyg) no proper motion could be derived.  

\subsection{Proper motion and binarity}

The angular separation between the positions of the C-star and the
maser reference component (given in Table\,\ref{tab1}) is 192\,mas
(142 and 130 mas in RA and Dec, respectively). The observed
difference seems to be significant ($\sim$ 4$\sigma$ above the
accuracy level from Table\,\ref{tab1} which is about 48\,mas), 
but the epochs of optical
and radio observations differ by about 10 years. The C-star would
need a proper motion of ($\mu_{\alpha}$\,cos(Dec), $\mu_{\delta}$)
close to ($-10$,$-10$) mas\,yr$^{-1}$ to produce the observed
difference in radio and optical positions.  
Khrutskaya et
al.\,(\cite{khr04}) have estimated the proper motion of V778\,Cyg from
the difference between the mean position of the star in the two epochs
of observations at the Pulkovo Observatory in (1935-1960) and in (1969-1980). The
Tycho2 catalog was used as a reference catalog for astrometric reduction of the
Pulkovo plates. Khrutskaya et al.\,(\cite{khr04}) estimate the proper
motion of V778\,Cyg to be in the range $\sim$[($-8.2 \pm -2.3$), 
($-3.9 \pm +8.4$)] 
mas\,yr$^{-1}$, not enough to match optical and radio positions 
even if observational errors are taken into account.

\begin{figure}[]
\resizebox{\hsize}{!}{\includegraphics{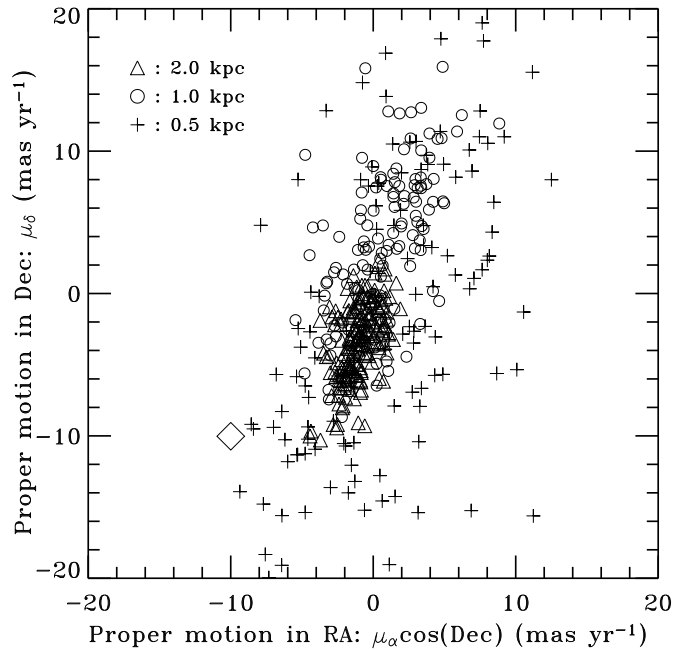}}
\caption{Different realisations of the proper motions for an evolved star with a LSR 
velocity close to $-$20\,km\,s$^{-1}$, which lies in the direction of 
V778\,Cyg at different distances (see text for details). The proper motion 
required to match optical C-star and radio maser position is marked by 
diamond.}
\label{Fig2}
\end{figure}

The problem of proper motion was further investigated using the
kinematic model of Groenewegen (2005, in preparation), which
includes Galactic rotation. For the present calculations, random
motions were drawn from Gaussian velocity ellipsoids with zero mean
and dispersions, typical for giants, of 31, 21 and 16 km s$^{-1}$
(Delhaye\,\cite{delhaye65}) in the $U,V,W$ directions, 
respectively\footnote{The following parameters also enter the model: a distance
Sun-Galactic Centre of 8.5 kpc (Kerr \& Lynden-Bell\,\cite{kerr86}),
Oort's constants of $A\,=\,14.4$ km\,s$^{-1}$\,kpc$^{-1}$,
$B\,=\,-12.0$ km\,s$^{-1}$\,kpc$^{-1}$, and higher order terms
$\frac{d^2\theta}{dr^2}_{|R_0}$\,=\,$-3.4$\,km\,s$^{-1}$\,kpc$^{-2}$,
$\frac{d^3\theta}{dr^3}_{|R_0}$\,=\,2.0\,km\,s$^{-1}$\,kpc$^{-3}$
(Pont et al.\cite{pontetal94}), and a Solar motion of 19.5 km s$^{-1}$
in the direction $l$\,=\,56\fdg0, $b$\,=\,23\fdg0 (Feast \&
Whitelock\,\cite{feast97}).}.
1000 simulations were performed to calculate proper motions and radial
velocities in the direction of V778 Cyg ($l$\,=\,95\fdg6,
$b$=\,+11\fdg51) for distances of 0.5, 1 and 2 kpc.  Fig.\ref{Fig2}
shows the results producing a $V_{\rm LSR}$ close to that of
V778\,Cyg, $-$20$\pm$5\,km\,s$^{-1}$, (determined from the
heliocentric radial velocity of about $-$35\,km\,s$^{-1}$, reported
by Barbier-Brossat \& Figon\,\cite{barbier00}). 
The distribution of points in Fig.\,\ref{Fig2} demonstrate that
it is unlikely  that an evolved star with 
$V_{\rm LSR}=$-$20\pm$5\,km\,s$^{-1}$ (V778\,Cyg) at a distance between 0.5
and 2.0 kpc would have a proper motion  sufficient to match the optical C-star 
position with the radio position. This would require a transverse velocity of 
order of ($-10$,$-10$) mas\,yr$^{-1}$, shown by the diamond in Fig.~2, which is 
outside all the predictions of our simulation. Note, that the  
distance to V778\,Cyg is estimated to be $D\simeq$1.4\,kpc (Peery et 
al.\,\cite{Peery75}; Yamamura et al.\,\cite{yamamuraetal00}).

Therefore, we believe that existing observations and simulations support a
binary system model discussed in detail by Yamamura et
al.\,(\cite{yamamuraetal00}) and that the water maser is associated with
a companion star. However, observational errors are not negligible
and simultaneous optical and radio observations would be required to 
confirm this hypothesis firmly.

The water maser components at $-$15, $-$17 and  $-$19 km\,s$^{-1}$
have been detected several times during the last
15 years (Nakada et al.\,\cite{nakada87}, Engels \&
Leinert\,\cite{engelsandleinert94}, this paper).  Comparison of these
observations show that changes in their radial velocities,
$\Delta$$V$, do not exceed 0.5\,km\,s$^{-1}$. If the velocity
change is due to the orbital motion of the secondary (with its maser)
around the carbon star (mass $M_{\rm c}$), the rate of change (independently 
of the companion mass) is given by: 
\begin{equation}
   \frac{\Delta V}{\Delta t} \simeq  \frac{G M_{\rm c}}{d^2} \sin i ,
\end{equation}
where $\Delta t$ is the time between observations, $G$ is the gravitational
constant and $i$ is the inclination angle. Hence, for $i\simeq$90\degr, the
distance $d$ between the binary companions is given by:
\begin{equation}
d \ga 75 \left( \frac{\Delta V}{0.5 {\rm km \; s^{-1}}}
\right)^{-1/2}
\left(\frac{\Delta t }{15 {\rm years} }\right)^{1/2}
\left(\frac{M_{\rm c}}{ M_{\odot}}\right)^{1/2}\,{\rm AU}.
\end{equation}

\subsection{Where do the water masers come from?}
\label{diskmodel}

\begin{figure}[]
\resizebox{\hsize}{!}{\includegraphics{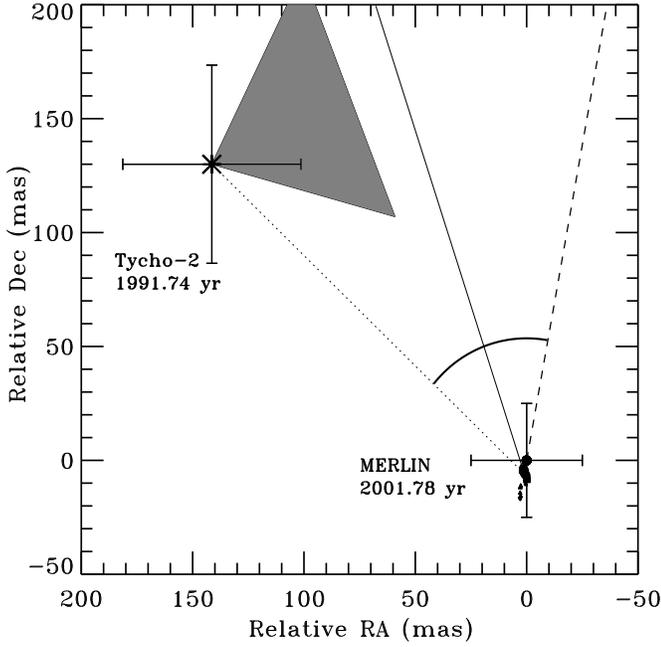}}
\caption{Relative positions (together with errors) of the 
C-star and water maser in V778\,Cyg. The assumed disk orientation is shown by
the straight solid line. The shaded region marks the allowed C-star locations
which are inferred from the proper motion estimations (see text for details). 
The sector of a circle marks the minimal allowed angular distance of the C-star
from the radio MERLIN position (as suggested by Eq.\,2) assuming that the 
distance to V778\,Cyg is 1.4\,kpc. The straight dotted line shows the position 
of orbital plane in case of no proper motion, while the dashed line marks the 
position angle of the overall structure seen in Fig.\,\ref{Fig1}.} 
\label{Fig3}
\end{figure} 
The distorted S--shaped structure seen in Fig.\,\ref{Fig1} can be
interpreted as a warped disk which may be non-co-planar 
with the orbital plane and is 
seen almost edge--on. The disk might be
made up of clumps which share the disk's rotation. Tidal forces 
(especially in a non-co-planar disk) could produce shear and/or
density waves (e.g. Papaloizou \& Terquem\,\cite{PT95}) responsible
for the observed distortion. We assume that the disk is at
P.A.=$+$18\degr, as suggested by the elongation angle of emission around
the central $V_{\rm LSR}\sim-$17\,km\,s$^{-1}$, and show the assumed 
orientation by the straight solid line in Fig.\,\ref{Fig3}. 
The shaded delimiting triangle, in this figure, marks the allowed C-star 
locations at the epoch of our observations (2001.78) as inferred from the 
proper motion estimations by Khrutskaya et al.\,(\cite{khr04}). The 
delimiting triangle has one corner at the optical C-star position (no proper 
motion at all) and the two others determined by the extreme limits to proper 
motion ($\mu_{\alpha}$\,cos(Dec), $\mu_{\delta}$) from Khrutskaya et 
al.\,(\cite{khr04}). The lower limit is derived from ($-8.2$,$-3.9$) and the 
upper (off the top of the plot) is derived from 
($-2.3$,$+8.4$)\,mas\,yr$^{-1}$. The allowed C-star locations determine
also the allowed positions (at the time of MERLIN observations) of 
the orbital plane of the system. The position of orbital plane in case of no 
proper motion is shown by the dotted line in Fig.\,\ref{Fig3}.

From Fig.\,\ref{Fig3} it is 
seen that the proposed interpretation is consistent with the observational 
constraints and the disk may be slightly (or may not be, if observational 
errors are taken into account) tilted relative to the orbital plane.
Note that assuming (as suggested by the overall structure seen in 
Fig.\,\ref{Fig1}) that the disk plane lies at position angle of
$-$10\degr\ (dashed line in Fig.\,\ref{Fig3}) we 
would not be able to fulfill constraints imposed by the proper motion 
estimation of V778\,Cyg (shaded region in Fig.\,\ref{Fig3}) by Khrutskaya et 
al.\,(\cite{khr04}), unless the disk is tilted by more than about 30\degr.

\begin{figure}[]
\resizebox{\hsize}{!}{\includegraphics{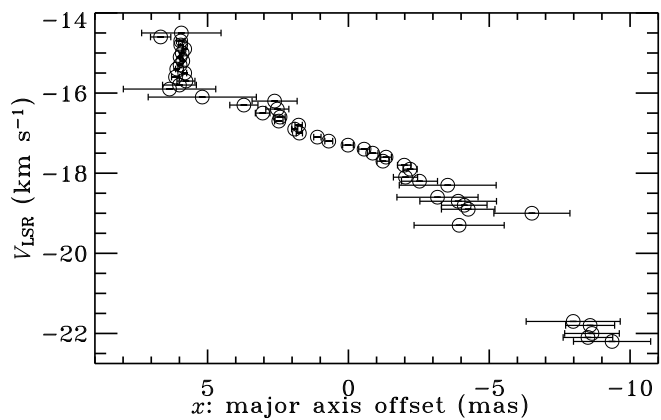}}
\caption{The LSR velocity against the distance along the major axis for water
maser components in V778\,Cyg. The major axis is assumed to be a line at 
P.A.=$+$18\degr\ defined by the $V_{\rm LSR}\sim-$17\,km\,s$^{-1}$ water maser
components; the zero of angular offset marks the centre of this structure. 
}
\label{Fig4}
\end{figure}

The above interpretation is strengthened by Fig.\,\ref{Fig4}, which shows the
$V_{\rm LSR}$ of the water maser components versus major axis offset $x$, 
($V-x$ diagram). The major axis is assumed to be at P.A.=$+$18\degr\ as
suggested by the elongation angle of emission around the central 
$V_{\rm LSR}\sim-$17\,km\,s$^{-1}$. Note, that we only detect one
unresolved maser component per channel and therefore the apparent
position of the components is along the line of sight of the greatest
amplification at the $V_{\rm LSR}$ sampled by that channel.
Fig.\,\ref{Fig4} shows an almost linear velocity gradient with a
greater velocity range per unit distance at the extremes.  This is
characteristic of emission from an almost edge--on disk in Keplerian
rotation (e.g. Shepherd \& Kurtz\,\cite{SK99}). As unsaturated
maser amplification is exponential, a relatively small change in
optical depth arising from a differential velocity gradient produces
strong domination by emission at the favored velocity (derived
rigorously in Pestalozzi et al.\,\cite{PECB04} and references
therein). In contrast, water masers associated with evolved-star jets
(Miranda et al.\,\cite{MGAT01}, Imai et al.\,\cite{IODOS02}) typically
show more fan--like structures with a less ordered velocity gradient
which tends to be steepest in the centre, not at the limbs. Note also
that the observed radial velocity gradient (about 7 km\,s$^{-1}$ in $\sim$17\,mas, 
i.e. about 0.3 km\,s$^{-1}$\,AU$^{-1}$ at distance of 1.4 kpc) seems to be too small 
for {\it typical} jets. In addition, the stability
of the maser velocities on the scale of 15 years (see Sect.\,3.1) 
suggests also that the maser originate from a stable configuration, but
not from an outflow or jet. Therefore we develop our discussion in the 
context of a disk model only. Multi-epoch (proper motion) observations 
are needed to rule out a jet completely.

Assuming that the disk is in Keplerian rotation and that the LSR
velocity of the disk center ($V^{\rm dc}_{\rm LSR}$) is about
$-$17\,km\,s$^{-1}$, we can express the line of sight velocity at
impact parameter $x$ (major axis offset) as $V_x$\,=\,$V^{\rm dc}_{\rm
LSR}+V_{\rm K}$\,$(x/R)$\,km\,s$^{-1}$, where $V_{\rm
K}$ is the assumed Keplerian velocity at the disk edge of radius $R$.  
The emission from an edge-on Keplerian disk has
three maxima: one corresponding to the systemic velocity (here about
$-$17\,km\,s$^{-1}$) and associated with matter on the line of sight
towards the disk centre (as modelled by Pestalozzi et
al.\,\cite{PECB04}), and two components (here, $-15$ and
$-19$\,km\,s$^{-1}$) associated with the outer edges of the disk where
emission peaks close to the Keplerian rotation velocity (see e.g.
Grinin \& Grigor'ev\cite{GG83}, Watson \& Wyld\,\cite{WW00}). For a
Keplerian velocity of 2\,km\,s$^{-1}$, the radius of the disk $R$ is
about $1/3$ larger than the impact parameter at which the amplification
is maximal $x_{\rm max}{\simeq}7$ mas.  This results in $R\,\simeq
\,(4/3)\times7\,$\,mas$\, \times$ $D$\,[kpc]\,=\,
13\,AU\,$(D/1.4)$\,kpc. Assuming that the mass of the disk itself
is negligible in comparison with mass of the secondary star, we can derive the central mass inside the disk using Kepler's laws, giving
$M_{\rm s}\simeq0.06\,(D/1.4\,{\rm kpc})\,{\rm M_\odot}$. Note, that
an object of such a small mass (brown dwarf?) would not be detected by
speckle interferometry as performed by Engels \&
Leinert\,(\cite{engelsandleinert94}).

The proposed model can easily explain components at $-$15, $-$17 and
$-$19\,km\,s$^{-1}$. The variations in their relative strengths could
be due to changes in physical conditions inside the disk: e.g. radial
temperature gradient (the temperature should be highest at the northern
disk edge), sub-sonic turbulence or clumpiness.  However, this model
cannot explain the weakest component at $-$22\,km\,s$^{-1}$. It is
possible that this feature is formed in material evaporating from the
disk (see model of Yamamura et al.\,\cite{yamamuraetal00}). On
the other hand, non--detection of an edge of the disk is also
possible. As exponential, directional maser amplification exaggerates
underlying conditions, some asymmetry is not surprising: there could
even be material at corresponding red--shifted velocities where masing
is temporarily disrupted. If we do not detect the extremities of
the disc, or if the angle of inclination is not quite 90\degr, the central 
star could be more massive (although still sub-Solar).
Finally, note that the observed strength of the water maser could be 
explained for the water concentration of about $10^{-5}$\,cm$^{-3}$ and the 
gas temperature in the disk of about 300\,K. Such a temperature is possible 
to achieve by the heating of dust inside the disk by the radiation from the 
C-star and the energy exchange between dust and gas (Babkovskaia et al. in 
preparation). 

\section{Conclusions}

We mapped the water maser emission from silicate carbon star V778\,Cyg
using MERLIN. The radio position obtained from MERLIN in 2001 is at an
angular separation of 190\,mas from the optical position of V778\,Cyg
given in the 1991 Tycho2 catalog. This cannot be explained by proper
motion if V778\,Cyg is at a distance of $\sim$1.4\,kpc and instead
(probably) provides observational support for the binary system model
of Yamamura et al.\,(\cite{yamamuraetal00}). Simultaneous radio and
optical measurements are needed to verify this model which will
be developed in a further paper. The velocity changes of the main
maser components over 15 years imply that the distance between the
C-star and disk is at least 75\,AU. The water maser components have an
almost linear, S-shaped distribution as projected on the sky and
in the variations of $V_{\rm LSR}$ along the elongation of the
main component at about $-$17\,km\,s$^{-1}$. We suggest that
this can be interpreted as an almost edge-on warped Keplerian disk located 
around a companion object and tilted by no more than 20\degr relative to the 
orbital plane. We estimate that the central mass inside the disk is 
$\ga$0.06\,M$_{\odot}$ (for a distance of 1.4\,kpc).

\begin{acknowledgements}
This work has  been supported by grant 2.P03D 017.25 of the Polish
State Committee for Scientific Research (RS),
 the Magnus Ehrnrooth Foundation, and the Finnish Graduate School for
Astronomy and Space Physics (NB), and the Academy of Finland (JP).

\end{acknowledgements}

\Online
\begin{table*}[]{}
\caption[]{Parameters of water maser components from MERLIN observations of V778\,Cyg.}
\label{tab2}
\begin{tabular}{l@{}c@{}l@{}l@{}l@{}l@{}c@{}}
\hline
channel & ~~V$_{LSR}$ (km\,s$^{-1})$ & ~~RA(J2000) & ~~Dec(J2000) & ~~Flux (Jy/beam) & ~~Flux error & ~~position error (mas)\\
\hline
103 & $-$14.5 & ~~20\,36\,07.38332 & ~~60\,05\,26.0238 & ~~3.8470E$-$01 & ~~2.60E$-$02 & 1.41 \\
104 & $-$14.6 & ~~20\,36\,07.38327 & ~~60\,05\,26.0247 & ~~1.5015E$+$00 & ~~2.61E$-$02 & 0.36 \\
105 & $-$14.7 & ~~20\,36\,07.38325 & ~~60\,05\,26.0240 & ~~3.5478E$+$00 & ~~2.68E$-$02 & 0.15 \\
106 & $-$14.8 & ~~20\,36\,07.38325 & ~~60\,05\,26.0240 & ~~7.1388E$+$00 & ~~2.57E$-$02 & 0.07 \\
107 & $-$14.9 & ~~20\,36\,07.38327 & ~~60\,05\,26.0238 & ~~1.2418E$+$01 & ~~2.89E$-$02 & 0.04 \\
108 & $-$15.0 & ~~20\,36\,07.38327 & ~~60\,05\,26.0239 & ~~1.9127E$+$01 & ~~3.18E$-$02 & 0.03 \\
109 & $-$15.1 & ~~20\,36\,07.38326 & ~~60\,05\,26.0240 & ~~2.4397E$+$01 & ~~2.49E$-$02 & 0.02 \\
110 & $-$15.2 & ~~20\,36\,07.38326 & ~~60\,05\,26.0239 & ~~2.1525E$+$01 & ~~3.30E$-$02 & 0.02 \\
111 & $-$15.3 & ~~20\,36\,07.38326 & ~~60\,05\,26.0240 & ~~1.9169E$+$01 & ~~3.06E$-$02 & 0.03 \\
112 & $-$15.4 & ~~20\,36\,07.38327 & ~~60\,05\,26.0241 & ~~1.3279E$+$01 & ~~2.93E$-$02 & 0.04 \\
113 & $-$15.5 & ~~20\,36\,07.38327 & ~~60\,05\,26.0238 & ~~7.6574E$+$00 & ~~2.69E$-$02 & 0.07 \\
114 & $-$15.6 & ~~20\,36\,07.38329 & ~~60\,05\,26.0241 & ~~3.9029E$+$00 & ~~2.63E$-$02 & 0.14 \\
115 & $-$15.7 & ~~20\,36\,07.38325 & ~~60\,05\,26.0238 & ~~1.7063E$+$00 & ~~2.60E$-$02 & 0.32 \\
116 & $-$15.8 & ~~20\,36\,07.38327 & ~~60\,05\,26.0240 & ~~8.9236E$-$01 & ~~2.63E$-$02 & 0.60 \\
117 & $-$15.9 & ~~20\,36\,07.38338 & ~~60\,05\,26.0241 & ~~3.3061E$-$01 & ~~2.68E$-$02 & 1.64 \\
118 & $-$16.1 & ~~20\,36\,07.38341 & ~~60\,05\,26.0228 & ~~2.8095E$-$01 & ~~2.54E$-$02 & 1.92 \\
119 & $-$16.2 & ~~20\,36\,07.38341 & ~~60\,05\,26.0201 & ~~6.7754E$-$01 & ~~2.68E$-$02 & 0.80 \\
120 & $-$16.3 & ~~20\,36\,07.38347 & ~~60\,05\,26.0211 & ~~1.0845E$+$00 & ~~2.63E$-$02 & 0.50 \\
121 & $-$16.4 & ~~20\,36\,07.38341 & ~~60\,05\,26.0200 & ~~1.3044E$+$00 & ~~2.63E$-$02 & 0.41 \\
122 & $-$16.5 & ~~20\,36\,07.38347 & ~~60\,05\,26.0204 & ~~2.0243E$+$00 & ~~2.56E$-$02 & 0.26 \\
123 & $-$16.6 & ~~20\,36\,07.38345 & ~~60\,05\,26.0198 & ~~3.0483E$+$00 & ~~2.69E$-$02 & 0.18 \\
124 & $-$16.7 & ~~20\,36\,07.38351 & ~~60\,05\,26.0197 & ~~3.4839E$+$00 & ~~2.54E$-$02 & 0.15 \\
125 & $-$16.8 & ~~20\,36\,07.38341 & ~~60\,05\,26.0192 & ~~4.2444E$+$00 & ~~2.63E$-$02 & 0.13 \\
126 & $-$16.9 & ~~20\,36\,07.38343 & ~~60\,05\,26.0193 & ~~5.0099E$+$00 & ~~2.68E$-$02 & 0.11 \\
127 & $-$17.0 & ~~20\,36\,07.38340 & ~~60\,05\,26.0192 & ~~4.7547E$+$00 & ~~2.67E$-$02 & 0.11 \\
128 & $-$17.1 & ~~20\,36\,07.38337 & ~~60\,05\,26.0186 & ~~4.2378E$+$00 & ~~2.67E$-$02 & 0.13 \\
129 & $-$17.2 & ~~20\,36\,07.38336 & ~~60\,05\,26.0182 & ~~3.7844E$+$00 & ~~2.73E$-$02 & 0.14 \\
130 & $-$17.3 & ~~20\,36\,07.38331 & ~~60\,05\,26.0176 & ~~2.8783E$+$00 & ~~2.62E$-$02 & 0.19 \\
131 & $-$17.4 & ~~20\,36\,07.38327 & ~~60\,05\,26.0171 & ~~2.9090E$+$00 & ~~2.51E$-$02 & 0.19 \\
132 & $-$17.5 & ~~20\,36\,07.38326 & ~~60\,05\,26.0168 & ~~2.8753E$+$00 & ~~2.66E$-$02 & 0.19 \\
133 & $-$17.6 & ~~20\,36\,07.38326 & ~~60\,05\,26.0163 & ~~2.8295E$+$00 & ~~2.56E$-$02 & 0.19 \\
134 & $-$17.7 & ~~20\,36\,07.38327 & ~~60\,05\,26.0164 & ~~2.4962E$+$00 & ~~2.68E$-$02 & 0.21 \\
135 & $-$17.8 & ~~20\,36\,07.38327 & ~~60\,05\,26.0156 & ~~2.2390E$+$00 & ~~2.64E$-$02 & 0.24 \\
136 & $-$17.9 & ~~20\,36\,07.38322 & ~~60\,05\,26.0155 & ~~2.1629E$+$00 & ~~2.66E$-$02 & 0.25 \\
137 & $-$18.1 & ~~20\,36\,07.38325 & ~~60\,05\,26.0156 & ~~1.2451E$+$00 & ~~2.65E$-$02 & 0.43 \\
138 & $-$18.2 & ~~20\,36\,07.38320 & ~~60\,05\,26.0152 & ~~8.3832E$-$01 & ~~2.59E$-$02 & 0.64 \\
139 & $-$18.3 & ~~20\,36\,07.38334 & ~~60\,05\,26.0138 & ~~3.1351E$-$01 & ~~2.65E$-$02 & 1.72 \\
141 & $-$18.5 & ~~20\,36\,07.38328 & ~~60\,05\,26.0156 & ~~1.7954E$-$01 & ~~2.60E$-$02 & 3.02 \\
142 & $-$18.6 & ~~20\,36\,07.38362 & ~~60\,05\,26.0135 & ~~3.7605E$-$01 & ~~2.62E$-$02 & 1.44 \\
143 & $-$18.7 & ~~20\,36\,07.38359 & ~~60\,05\,26.0128 & ~~3.9566E$-$01 & ~~2.65E$-$02 & 1.36 \\
144 & $-$18.8 & ~~20\,36\,07.38366 & ~~60\,05\,26.0124 & ~~6.6483E$-$01 & ~~2.58E$-$02 & 0.81 \\
145 & $-$18.9 & ~~20\,36\,07.38364 & ~~60\,05\,26.0123 & ~~5.6821E$-$01 & ~~2.55E$-$02 & 0.95 \\
146 & $-$19.0 & ~~20\,36\,07.38365 & ~~60\,05\,26.0099 & ~~4.0059E$-$01 & ~~2.64E$-$02 & 1.35 \\
147 & $-$19.1 & ~~20\,36\,07.38363 & ~~60\,05\,26.0115 & ~~2.1827E$-$01 & ~~2.58E$-$02 & 2.47 \\
148 & $-$19.2 & ~~20\,36\,07.38317 & ~~60\,05\,26.0121 & ~~1.5076E$-$01 & ~~2.61E$-$02 & 3.60 \\
149 & $-$19.3 & ~~20\,36\,07.38329 & ~~60\,05\,26.0135 & ~~3.2427E$-$01 & ~~2.55E$-$02 & 1.66 \\
172 & $-$21.7 & ~~20\,36\,07.38363 & ~~60\,05\,26.0084 & ~~3.2280E$-$01 & ~~2.50E$-$02 & 1.67 \\
173 & $-$21.8 & ~~20\,36\,07.38366 & ~~60\,05\,26.0077 & ~~6.2120E$-$01 & ~~2.66E$-$02 & 0.87 \\
174 & $-$22.0 & ~~20\,36\,07.38367 & ~~60\,05\,26.0076 & ~~5.5633E$-$01 & ~~2.70E$-$02 & 0.97 \\
175 & $-$22.1 & ~~20\,36\,07.38361 & ~~60\,05\,26.0079 & ~~6.0993E$-$01 & ~~2.56E$-$02 & 0.88 \\
176 & $-$22.2 & ~~20\,36\,07.38365 & ~~60\,05\,26.0069 & ~~3.9179E$-$01 & ~~2.70E$-$02 & 1.37 \\
\hline
\end{tabular}
\medskip
\begin{list}{}{}
\item[The table provides the spectral channel number (Col.\,1), the velocity (in respect to 
the local standard of]
\item[rest) of the detected water maser components (Col.\,2), the absolute 
coordinates (RA and DEC at J2000)]
\item[of each feature (Cols.\,3 and 4, respectively), flux of 
each maser component and the corresponding flux]
\item[error (Cols.\,5 and 6, respectively), and 
the position uncertainty of each component (Col.\,7).]
\end{list}
\end{table*}
\end{document}